# Computational interference microscopy enabled by deep learning

YUHENG JIAO,[1,2] YUCHEN R. HE,[1] MIKHAIL E. KANDEL,[1] XIAOJUN LIU,[2] WENLONG LU,[2] AND GABRIEL POPESCU[1, *]

[1]*Quantitative Light Imaging Laboratory, Department of Electrical and Computer Engineering, Beckman Institute for Advanced Science and Technology, University of Illinois at Urbana-Champaign, Illinois 61801, USA*
[2]*State Key Laboratory of Digital Manufacturing Equipment and Technology, School of Mechanical and Engineering, Huazhong University of Science and Technology, Wuhan 430074, China*
[*]*gpopescu@illinois.edu*

**Abstract:** Quantitative phase imaging (QPI) has been widely applied in characterizing cells and tissues. Spatial light interference microscopy (SLIM) is a highly sensitive QPI method, due to its partially coherent illumination and common path interferometry geometry. However, SLIM's acquisition rate is limited because of the four-frame phase-shifting scheme. On the other hand, off-axis methods like diffraction phase microscopy (DPM), allows for single-shot QPI. However, the laser-based DPM system is plagued by spatial noise due to speckles and multiple reflections. In a parallel development, deep learning was proven valuable in the field of bioimaging, especially due to its ability to translate one form of contrast into another. Here, we propose using deep learning to produce synthetic, SLIM-quality, high-sensitivity phase maps from DPM, single-shot images as input. We used an inverted microscope with its two ports connected to the DPM and SLIM modules, such that we have access to the two types of images on the same field of view. We constructed a deep learning model based on U-net and trained on over 1,000 pairs of DPM and SLIM images. The model learned to remove the speckles in laser DPM and overcame the background phase noise in both the test set and new data. Furthermore, we implemented the neural network inference into the live acquisition software, which now allows a DPM user to observe in real-time an extremely low-noise phase image. We demonstrated this principle of computational interference microscopy (CIM) imaging using blood smears, as they contain both erythrocytes and leukocytes, in static and dynamic conditions.

## 1. Introduction

Quantitative phase imaging (QPI) has developed into an active field with the goal of providing a label-free alternative to biomedical imaging, complementary to the standard techniques relying on stains and fluorescent tags.[1] QPI yields the optical pathlength map associated with the specimen and, thus, informs about both the thickness and the refractive index of the structure of interest. Due to its quantitative and nondestructive nature, QPI has found important biomedical applications ranging from basic science to clinical diagnosis.[2] As the specimen refractive index reports on the dry mass density, QPI has been employed to studying cell growth [3-7]. Analyzing the spatio-temporal fluctuations of dry mass provided a new way of monitoring intracellular transport and differentiating between diffusive and active processes [8-10]. Due to its sensitivity to nanometer scale optical pathlength changes, QPI is capable of measuring cell membrane fluctuations [11-14] and imaging unlabeled single microtubules [15]. The full holographic information (phase and amplitude) associated with a field scattered by a transparent object allows for tomographic reconstructions without ambiguity, as demonstrated by Wolf in 1969[16]. Thus, QPI-based tomography has been demonstrated by acquiring phase imaging data as a function of illumination angles [17-19], scanning the object through focus

[20, 21], and performing spectroscopic measurements [22]. This approach has been recently extended to second harmonic fields [23]. QPI has led to the discovery of new intrinsic markers for cancer diagnosis and prognosis, without the variability generally introduced by stains [24-31]. More recently, QPI has been extended to strongly scattering specimens, such as embryos, spheroids, and acute brain slices [32, 33], which expanded significantly QPI's range of applications.

In terms of methodology, QPI instruments are divided into off-axis [34-38] and phase shifting [32, 39, 40] geometries. Phase shifting methods (see Chapter 10, in Ref. [1]) use time domain modulation and, as a result, preserve the maximum space-bandwidth product for a given imaging instrument, at the expense of frame rate. Conversely, off-axis interferometry (see Chapter 9 in Ref. [1]) operates on spatial modulation, which provides single shot, fast imaging, at the expense of space-bandwidth product. In addition, due to the spatial domain image processing, off-axis methods also generally result in higher spatial phase noise. This spatial noise is further amplified by speckles whenever monochromatic rather than broadband light is used [see Ref [39] for an assessment of the spatial noise in both geometries].

An ideal QPI method would provide the low noise, high resolution associated with phase shifting interferometry and single-shot performance associated with off-axis geometries. In this paper, we demonstrate that such performance can be achieved by using deep learning to produce an image-to-image translation from single shot noisy data to phase-shifting, low-noise images, on which the network was a priori trained. For generating the input data, we used diffraction phase microscopy [34], an off-axis, common path QPI technique, while for ground truth, we used images of the same field of view obtained by spatial light interference microscopy [39], which is a broadband light, phase-shifting, common path method. A U-Net convolutional neural network was trained to infer a SLIM image from a DPM image as input. The performance of this image translation was significant, with a peak signal-to-noise ratio (PSNR) around 30 and a Pearson correlation coefficient of 0.79. Finally, we integrated the inference algorithm into the acquisition software, which allows for real-time operation.

## 2. Results

*2.1 DPM and SLIM data collection*

In order to acquire training data necessary to produce SLIM-quality images in a single-shot, we developed a combined SLIM-DPM system, which generates both images from the same field of view (Figure 1). The DPM and SLIM modules were placed at the two side ports of a commercial inverted microscope (Axio Observer Z1, Zeiss). A coupled fiber green laser ($\lambda = 532\ nm$) was used as illumination for DPM, with the condenser aperture closed to minimum. A collimated LED source ($\lambda = 623\ nm$, 20 nm bandwidth) was used as illumination for SLIM (CellVista SLIM Pro, Phi Optics, Inc.), with the conventional ring illumination associated with phase contrast microscopy. Using 20x/0.4 NA objective, the magnified image is replicated to either port by using a switch.

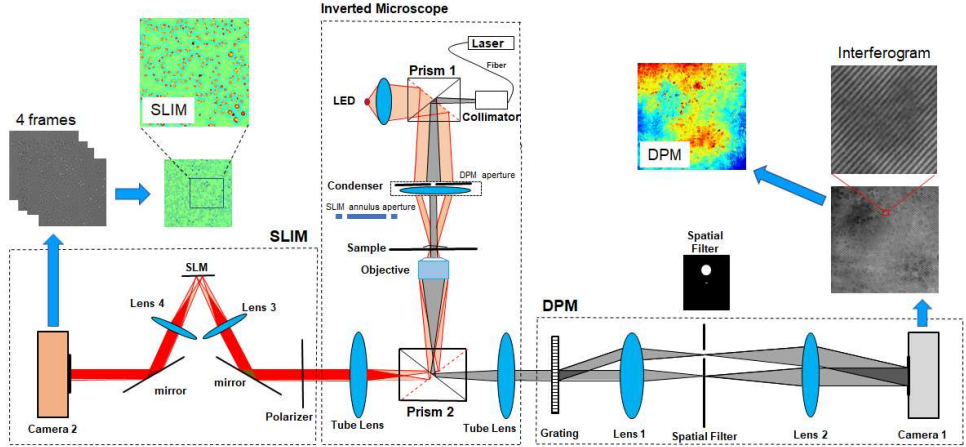

Fig. 1. Schematic of the imaging setup. The system is built around an inverted microscope. We are using a 20x/0.4NA objective. The two side ports connect to the DPM (right) and SLIM (left) modules. Thus, we obtain SLIM and DPM images on the same field of view. The focal length of lenses 1 and 2 are 100 mm and 200 mm, respectively. Lenses 3 and 4 have the same focal length. When switching between DPM and SLIM, Prisms 1 and 2 are switched to different positions and the condenser is set to PH1 set. Due to the magnification of the 4f system in DPM, a registration is needed to match the DPM and SLIM images.

The DPM principle has been presented in detail previously [41]. Briefly, a phase diffraction grating is placed at the image plane to replicate the spatial frequency content at the Fourier plane of Lens 1, along multiple diffraction orders. The 0th order is spatially filtered to remove all higher frequencies, the 1st order beam is passed unaltered, and all the remaining orders are completely blocked. Thus, following lens 2, the camera detects the interference between the image field and the filtered 0th order, which acts as a reference field. There is a further 2.5X magnification produced by the L1-L2 4-f system. The resulting interferogram is processed using the well-known Hilbert transform to retrieve the phase map from a single recording. The exposure time for each DPM image was 2 ms throughout all our measurements. As illustrated in Fig.1, the resulting DPM image is affected by background nonuniformities, which in the past have been mitigated using post processing [42].

The SLIM module (see Ref [39] for details) contains a phase-only spatial light modulator (SLM) at the pupil plane created by lens 3. A binary mask is created on the SLM to precisely match the phase contrast ring placed in the condenser aperture. By shifting the phase of the mask in increments of $\pi/2$, we record 4 intensity images using Camera 2. The four frames are combined to obtain the quantitative phase map, as described in Ref [39].

To validate our phase measurements, we performed a side-by-side comparison of the two measurements on a 1 $\mu m$ polystyrene bead, immersed in oil (Zeiss, Immersol 518F), as illustrated in Fig. 2. The refractive index of the beads is 1.588 at 623 nm wavelength, while that of the oil is 1.518. As a result, the expected phase shift from the bead using red led is

$$\phi = 2\pi \, \mathrm{d}\,(n - n_0)/\lambda = 0.71 \,.$$

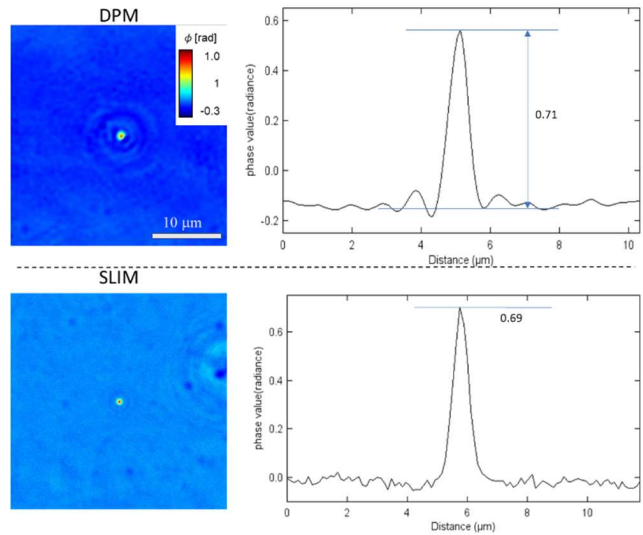

Fig. 2. Oil-immersed, 1 μm diameter polystyrene beads imaged by DPM and SLIM, as indicated. The measured phase values are comparable to the expected number, as detailed in text.

For comparison, the DPM phase map, which was imaged with a green laser, was normalized by the ratio of two wavelengths. The values obtained by DPM and SLIM, 0.71 rad and 0.69 rad are compatible with the expected value of 0.71 rad.

*2.2. Training procedure*

In order to use deep learning to infer SLIM images from DPM data, we employed the U-Net described in Figure 3. Our model was a modified version of the U-Net [43]. We added residual connection within each feature-extracting block in the encoder path and batch normalization layers [44] between convolution and activation layers. Three drop-out layers were added in the encoder path and the bottleneck to avoid overfitting. Our model had far fewer trainable parameters than the original U-Net, as we reduced the number of channels in each layer to a quarter of the original proposed value. The model was optimized using adaptive moment estimation (Adam) [45] against the mean-squared error. All the input values were scaled into the range of [0, 1].

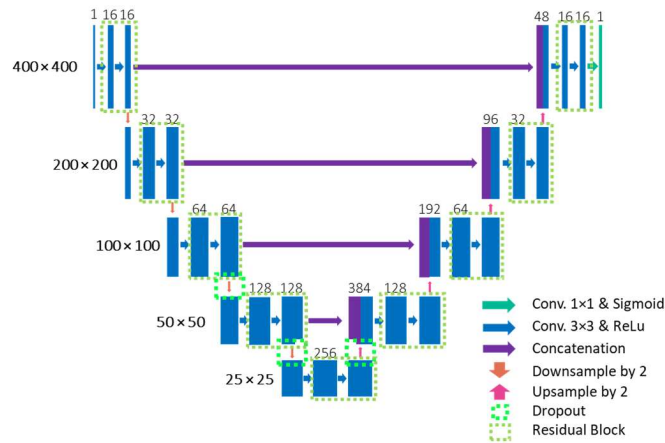

Fig. 3. The U-net structure. The model has a symmetric layout and consists of three major parts: the encoder path, the bottleneck, and the decoder path. We added Batch Normalization layers between convolution and activation layers to stabilize the learning process. After every 2 convolutional operations, a residual connection was added for faster convergence. 3 drop-out layers were also used to avoid overfitting. The kernel size was set to 3×3 and the number of kernels was reduced to a quarter of the original proposed number. A 400×400 region was randomly cropped from each original 1536×1536 image during training.

To demonstrate the deep-learning-enabled computational interference microscopy (CIM) operation on live cells, we used blood cell smears, which contain red blood cells and several types of white blood cells. More than 1,200 images were recorded by both SLIM and DPM, with over 100 cells in each field of view. To boost variation within our dataset, the images were cropped into random 400 × 400 pixel regions and fed into the U-net for training. The network was trained for 21 hours on a single GTX 1070 GPU, using 1,000 epochs.

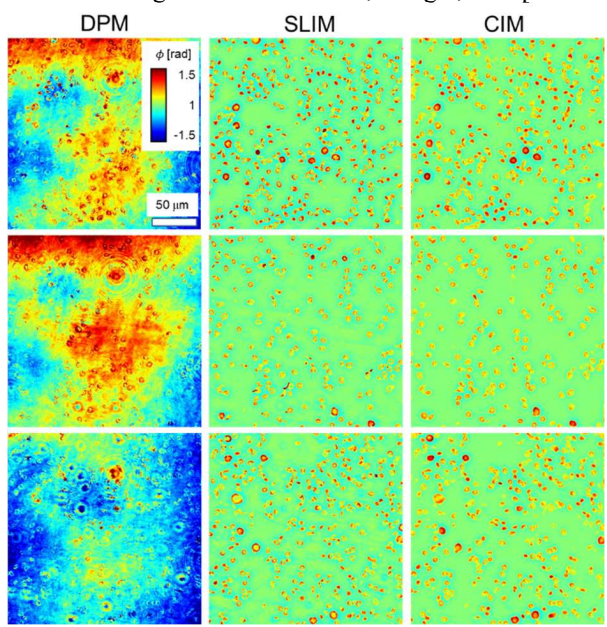

Fig. 4. DPM input data (left column), SLIM ground truth data (center) and the U-net inference (right). All images share the same calibration and scale bar. As can be seen, the neural network correctly infers SLIM images from the DPM input, with drastically reduced noise levels.

Figure 4 illustrates the DPM input data (left column), SLIM ground truth (middle), and the resulting CIM (right). Visually, the U-Net is able to reduce the overall noise of the DPM input and produce a remarkably similar image to the SLIM ground truth. In order to quantify the performance of the neural network, we computed both the peak signal to noise ratio (PSNR) and the Pearson correlation coefficient between the ground truth and the prediction. The mean-squared loss value on the training dataset and the validation dataset after each epoch is plotted in Figure 5. The model checkpoint with lowest loss value on the validation dataset was selected as our end model for evaluation and deployment. Figure 5 also shows the result of the PSNR and Pearson correlation conducted on the training, validation, and test datasets. The results are consistent among all three datasets, indicating that the model generalized well on the unseen test dataset.

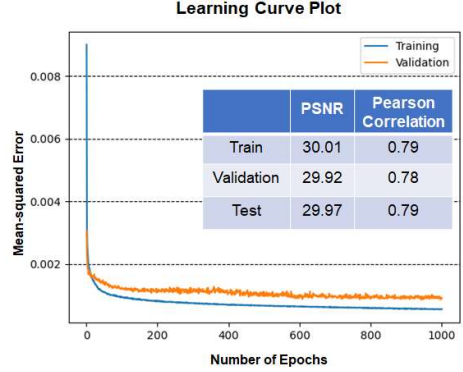

Fig. 5. Model convergence and performance. After each epoch, we plotted the mean-squared loss value on both the training dataset and validation dataset. The model checkpoint with the lowest error on the validation dataset was selected as our end model for evaluation and deployment. The average PSNR and Pearson correlation coefficient of our model on the training, validation, and test datasets were computed. Our models generalized well on the unseen test dataset.

## 2.3 Real-time inference

To boost the usability of CIM, we integrated the inference algorithm into the real-time acquisition software for DPM (wDPM CellVista Pro, Phi Optics, Inc.). Figure 6 shows the user interface, with the DPM image being reconstructed and used as input for inference. Supplemental video Visualization 1 illustrates the real-time operation. The conversion from the noisy DPM images to CIM takes place at a push of a button. Note that translating the stage does not affect the quality of the inference, which works very well for both red and white blood cells. Thus, we envision that CIM can be readily used for automating large field of view and multi-well plate scanning. To integrate the trained model into our acquisition software, we built the same network architecture from scratch using the Nvidia TensorRT library and loaded the transposed trained weights. The computer was equipped with a Nvidia RTX2070 GPU. The software then ran the model inference on a 1536 × 1536 image at a speed of 58 ms per image. Thus, the resulting CIM is computed and displayed in real time at a rate of up to 14 frames per second. This performance can be boosted further by either lowering the pixel size or upgrading the GPU.

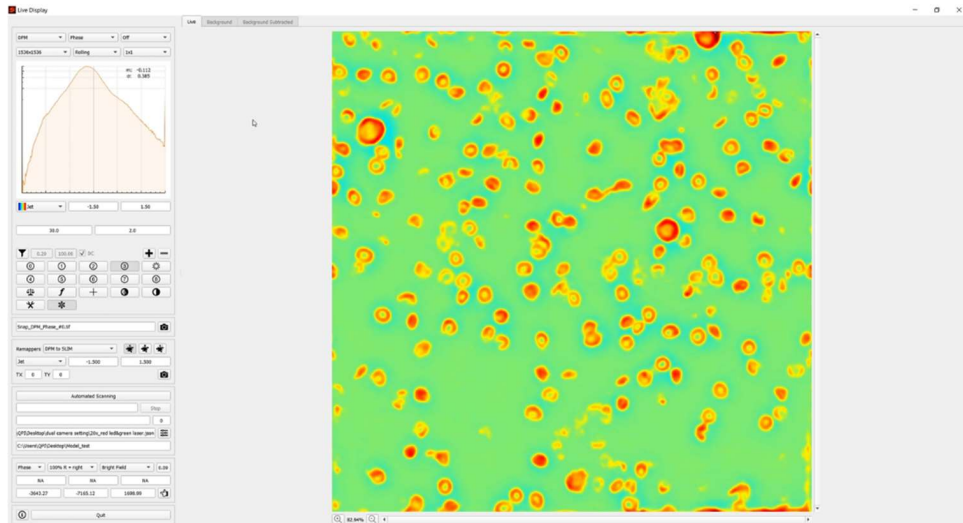

Fig. 6. Live imaging interface. We demonstrate our live imaging system through the three channels: DPM, SLIM, and CIM channel (see Visualization 1). The image shows the graphic user interface and a snapshot of CIM operation in real-time.

Supplemental video Visualization 2 shows a CIM time-lapse of a fresh, unlabeled blood smear. We collected blood from a healthy volunteer, diluted one drop of blood with 10 ml Phosphate-buffered saline. No effect was devoted to stabilizing the smear, such that we can test the ability of CIM to operate on highly dynamic samples. These specimens were never imaged by SLIM, thus, the network was not trained on these samples. Clearly CIM performs very well and even reveals minute membrane fluctuation in individual red blood cells. The area of cell overlapping sometimes display lower phase values than expected, but this appears to be an optical, rather than computational artifact, as it is also present in Fig. 4 (SLIM column). These data highlight the capability of CIM to run under flow conditions, for applications such as flow cytometry. Figure 7 shows 4 sequential frames from the time lapse in Visualization 2, taken 100 ms apart, with 2 ms exposure each. The cells selected in the rectangular boxes moved very fast, as can be visualized in the movie and these snapshots. These results prove that the CIM system can provide single-shot, high speed measurements, as allowed by DPM, while the output has the high quality of SLIM images.

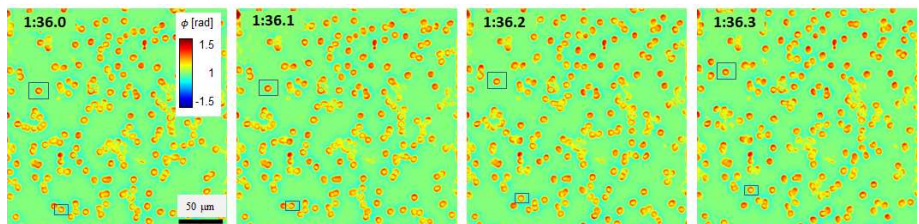

Fig. 7. Dynamic imaging of blood cells with snapshots at different moments in time, 100 ms apart, as indicated. The scale bar and colorbar are the same for all the frames. The full video can be visualized as Visualization 2. In these frames, one can see cells (e.g., those in the rectangular boxes) that flow very fast. However, the CIM inference is accurate and operates in real-time.

## 3. Methods

*3.1 Deep Learning*

We formulated the problem as an image-to-image regression, where the deep neural network takes in a DPM image as input and predicts a new image that is close to a SLIM image of the same field of view. Our neural network was a variant of U-Net, which has shown great performance on similar tasks with quantitative phase imaging data [46]. Unlike the stock U-Net, we added Batch Normalization layers between convolution and activation layers to stabilize the learning process. We also added in a residual connection after every two convolution operations (on the same field of view) for faster convergence and better performance. The model has a symmetric layout and consists of three major parts: an encoder path, a bottleneck and a decoder path. The encoder path captures contextual information in the image. It consists of 4 stages of convolutional and non-linear activation operations with residual connection. Each stage was followed by a 2×2 downsampling operation. The decoder path is almost symmetric to the encoder path, except that it has upsampling operations to combine low-resolution and high-resolution information and enables localization. The convolution kernel size within the network was set to 3×3 except for those used in residual connection, which was set to 1×1. The number of kernels in each stage were set to 16, 32, 64, and 128 respectively, reduced to a quarter of the proposed value in the original U-Net paper [43]. Thus, our model had only 3.3 million trainable parameters. Based on the training results, it was apparent that this model was already complex enough to approximate the transform from DPM images to SLIM images. We picked the mean-squared error as our loss function and used the Adam optimizer with the default exponential decay rate for both moment estimates (0.9 and 0.999, respectively). We used PSNR and Pearson correlation coefficient to measure the performance of our network. All the input images (DPM and SLIM) were scaled from $[-\pi, \pi]$ to $[0,1]$. We trained the model from scratch with a learning rate of 6e-5 for 1,000 epochs. The batch size was set to 4 during training. The mean squared loss value on both the training and the validation dataset was plotted after each epoch. To add more variation during the training process, we applied random cropping to each training image. A 400 x 400 crop was selected randomly from each original 1536 x 1536 image and fed into the model during one epoch. Since U-Net is fully convolutional, it can pick up features on these smaller crops and apply them later onto the larger images. This random cropping has two main advantages. First, it served as a form of data augmentation, contributing to better generalizability. Second, it reduced the training time and GPU memory requirement. The model was implemented using TensorFlow and the training was performed on a GTX 1070 GPU with 8GB memory. The training took approximately 21 hours.

*3.2 Real-time implementation*

To incorporate the trained model into our acquisition software, we first saved all the trained weights (convolution kernels and batch normalization parameters) into a single HDF5 file. Then we constructed the same network architecture within our acquisition software using the Nvidia TensorRT API in C++. Due to the mismatch between weight formats in TensorFlow and TensorRT, the weights were first transposed and then loaded into the network architecture in C++. Once the network was built, TensorRT optimized the inference procedure by enumerating different configuration of kernels. This optimization was necessary because the optimal configuration for inference differ on from hardware to hardware. Thus, we overlapped the model optimization with the software initialization by utilizing multi-threading in C++.

## 4. Discussion

QPI has developed into several types of methods, characterized by different advantages and drawbacks according to the specific geometry. The laser DPM enables high acquisition rate, but is plagued with speckles and background phase noise. White light DPM overcomes the noise with broadband light at the expense of intensity loss, thus, a lower acquisition rate. Because of illumination NA, SLIM has a higher transverse resolution than the DPM based methods. However, the acquisition speed of SLIM is further limited by the four phase shifts.

Artificial intelligence has already demonstrated its potential to transform biomedical imaging. In microscopy, image-to-image translation techniques proved valuable in generating synthetic fluorescence images for cell biology applications [47, 48] and digital staining for pathology applications [49]. Phase imaging with computational specificity (PICS) has been introduced recently to retrieve molecular specificity via deep learning, thus, eliminating the phototoxicity and photobleaching associated with fluorescence tags[50] and intrinsic fertility makers [51].

Here, we presented a different application of AI, where the goal was to combine the benefits of different QPI instruments. CIM is a new QPI method that combines a high acquisition rate from DPM and low-noise and high resolution from SLIM. This approach allows for applications that involve both highly dynamic and sensitive measurements. Since phase is the intrinsic marker of transparent samples, we anticipate that our current model can be extended to other specimens via transfer learning with a small amount of new data, making our method extendible to new biological samples at low costs.

## Acknowledgements

We are grateful to the following sources of funding: National Institute of General Medical Sciences (GM129709); National Cancer Institute (CA238191); National Science Foundation (2013AA014402, CBET0939511 STC, NRT-UtB 173525). Yuheng Jiao was supported by CSC: 201806160045; MEK is supported by MBM (NSF, NRT-UtB, 1735252).

## Disclosures

G.P. founded Phi Optics, Inc., a company that commercializes QPI methods for biomedical applications.